# A Context-based Trust Management Model for Pervasive Computing Systems

Negin Razavi
*Islamic Azad University, Science and Research Branch*
*Tehran, Iran*

Amir Masoud Rahmani
*Islamic Azad University, Science and Research Branch*
*Tehran, Iran*

Mehran Mohsenzadeh
*Islamic Azad University, Science and Research Branch*
*Tehran, Iran*

*Abstract*—Trust plays an important role in making collaborative decisions about service evaluation and service selection in pervasive computing. Context is a fundamental concept in pervasive systems, which is based on the interpretation of environment and systems. The dynamic nature of context can strongly affect trust management and service selection. In this paper, we present a context-based trust management model for pervasive computing systems. The concept of context is considered in basic components of the model such as trust computation module, recommender assessment module, transaction management module, and request responder. In order to measure a predicted trustworthiness according to the fuzzy nature of trust in pervasive environments, fuzzy concepts are integrated in the proposed model.

I. INTRODUCTION

Having an effective trust management model plays an important role in evaluating relationships among communicating entities. Communicating entities in pervasive computing environments may include service requesters and service providers. Pervasive computing is an emerging research field that improves revolutionary paradigms for computing models. An efficient pervasive computing model depends on the trust management model. A pervasive environment is consisting of multi-resources (or multi service providers) which demonstrate the need for an efficient model for trust management. The fundamental issue in pervasive computing is to create ambient intelligence where embedded entities in the environment provide persistent connectivity and service without awareness of communications or computing technologies.

In pervasive computing environment, having efficient and trusted relationships and privacy solutions both together is a challenge [1]. Besides pervasive computing prepares additional features and functionality, such as invisibility and context awareness [2]. As a result, presenting a trust management model can help us to have a safely structure for our pervasive environment.

Trust management concept was used in 1996 for the first time to solve network problems [3]. In [4] a research is done on trust management models in pervasive computing and a model is suggested to facilitate the interactive of entity and environment and improve the efficiency of trust-building. Some approaches use special techniques in their trust models. In [5] a model based on cloud theory is presented. This theory is used to describe uncertain concepts such as trust. The model proposed in [6] uses a new feature of gravitation to analyze the trust relationship between pervasive interaction entities.

Performance and privacy protection are other concepts which are considered in recent researches. [7] presents a privacy-preserving credential chain discovery mechanism for credential chain discovery problem in trust management. By this mechanism, credentials are no longer available to everyone. In [8] a specific framework is presented for implementing the distributed trust scheme and the performance is evaluated for the metrics of throughput, packet loss ratio and message overhead.

Most decisions about establishing relationships between entities or selecting services among different service providers in a pervasive environment depend on the concept of context. A context relates information types with resources in the environment, and provides a Situation Derivation Function that gathers actual information [9] which can influence the interactive behavior of entities. The meaning of the term context may vary, dependent on the system or the domain of usage. Distance, packet rate, packet loss ratio, time, and delay are some examples of context in different domains. For more efficiency, information gain from context can be used in trust management model in order to help decision making.

In this paper, we propose a context-based trust management model which increases the reliability of interactions such as service selection. In the proposed model, most of the components rely on context to obtain effective functionality. Recommendations from other entities are used in order to achieve faster and more accurate trust evaluation. As mentioned in [10], trustworthiness measurement and prediction are complex and limited by the fuzzy, dynamic and complex nature of trust. Therefore, we also consider fuzzy concepts in our trust management model. Privacy protection has always been the subject of legislation, since there is an inherent conflict in service provisioning [11]. We use a privacy agent for satisfying privacy-levels in our model.

None of the previous works considers the concept of context and its fundamental role in the trust management. The advantage of our work is that the context concept is considered in the proposed trust management model.

*Outline of the paper:* In Section II, an overview of basic concepts in the proposed trust management model is presented. Section III defines trust management model at service requester and Section IV defines trust management model at service provider. Section V describes the procedure of trust computation in the proposed model. Some characteristics of the proposed model are represented in Section VI and finally the conclusion and future works are given in Section VII.

## II. AN OVERVIEW OF BASIC CONCEPTS IN THE PROPOSED TRUST MANAGEMENT MODEL

### A. Definition of Trust

Trust has been defined in several ways. In our trust management model, the trust of entity $A$ on entity $B$ shows the strength of $A$'s belief that $B$ can provide a service, which will satisfy $A$'s request, and that the behavior of $B$ is without malicious intent. This definition is based on the definition of trust in [12], [13].

In this model, trust values (TRV) range from -1 to 1. $TRV=0$ indicates that the service requester has no trust information about the service provider, $TRV > 0$ indicates that the service provider is considered trustworthy, and $TRV < 0$ indicates that the service provider is considered untrustworthy.

### B. Definition of Context

As mentioned in [14], context is any evidence that can be used to support arguments for the conditions of the situation of any entity or target, which influences their interactive behavior. Privacy, security, and trust may be representatives of the rules that influence the interactive behavior between entities. Therefore, the concept of context is integrated with other concepts such as privacy, security, and trust.

### C. Definition of Service

In our model, entities interact with each other by means of services. A service can be presented with an array of attributes where different attributes in services result in different service types. Each service type has one (or more) critical context(s) that can affect the selection of the target entity or influence the provided quality of service (QoS). Time, delay, and distance can be treated as context. When several entities afford the same kind of services, the service requester needs to handle critical contexts while services are provided.

### D. Model Framework Analysis

In the proposed model, we define two types of entities: service requester and service provider. As shown in Fig. 1,

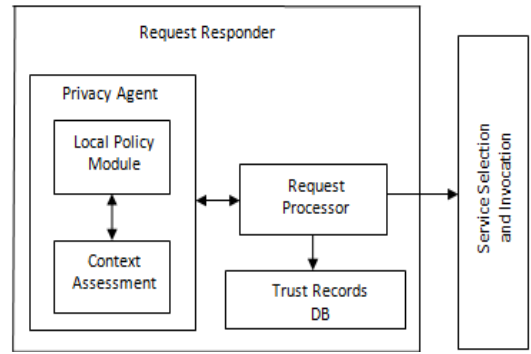

Figure 1. Trust management model at service requester

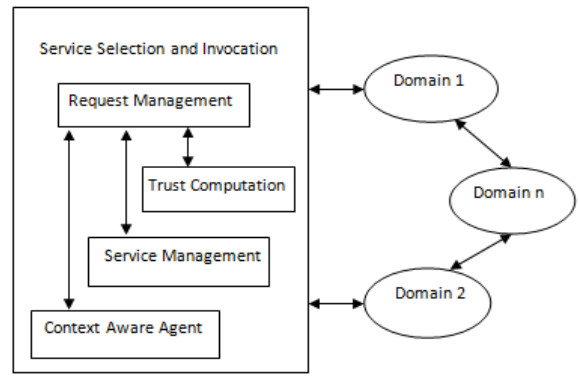

Figure 2. Trust management model at service provider

each service requester has a component, named *service selection and invocation*, which interacts with domains of the environment. Each service provider has a component named *request responder*, shown in Fig. 2, which is responsible for the request arrivals. *Service selection and invocation* and *request responder* are the two major functional components in the framework.

## III. TRUST MANAGEMENT MODEL AT SERVICE REQUESTER

*Service selection and invocation* component is composed of *service management* module, *request management* module, and *trust computation* module at service requester. The functionalities of these modules together result in service request and selection by the corresponding component. *Service selection and invocation* component uses a *context aware agent* for the purpose of context maintenance.

CONTEXT AWARE AGENT: This agent divides the environment into domains, considering the contexts which are critical in service selection. As a result, each domain contains entities that are suitable for a type of service. For example, in mobile services, this agent can create domains in which the distance of the entities from the requester is not more than 1 km.

| $Att_1$ | $Att_2$ | ... | $Att_n$ |
|---------|---------|-----|---------|
| $TV_1$  | $TV_2$  | ... | $TV_n$  |

Figure 3. Representation of a service request

SERVICE MANAGEMENT MODULE: This module identifies the required service type and denotes the threshold value (*TV*) of each service attribute which is required for a service to be satisfied. *Service management* module also sends requests to the selected service provider which is determined by the *request management* module.

Fig. 3 represents a service request. $Att_i$ denotes $i^{th}$ service attribute and $TV_i$ denotes the threshold value of $Att_i$.

REQUEST MANAGEMENT MODULE: After determining the service type and required threshold values by *service management* module, the request is passed to *request management* module. This module sends the request to the domains that are identified by *context aware agent*. After that, entities which can provide the requested service, send their real values to *request management* module. *Request management* module uses a fuzzy evaluation to select the best service provider and informs *service management* module to send the service request to the selected service provider. If none of the entities in the corresponding domains respond, *request management* module broadcasts the request to all the domains. The fuzzy evaluation function of service providers is as following.

$$SP_i = W \cdot M_i = (w_1, w_2, ..., w_n) \cdot \begin{pmatrix} m_{i_{11}} & m_{i_{12}} & m_{i_{13}} \\ m_{i_{21}} & m_{i_{22}} & m_{i_{23}} \\ . & . & . \\ . & . & . \\ . & . & . \\ m_{i_{n1}} & m_{i_{n2}} & m_{i_{n3}} \end{pmatrix}. \quad (1)$$

$$\sum w_j = 1; \; 1 \leq j \leq n; \; 1 \leq i \leq k$$

Here, *n* is the number of service attributes and *k* is the number of service providers. Element $w_j$ in the array *W* represents the weight factor for the $j^{th}$ service attribute which reflects the importance of the attribute. Element $m_i$ in matrix *M* represents the membership degree for real values proposed by $i^{th}$ service provider with respect to the quality level that can be *Good*, *Average* and *Bad*. $SP_i$ is the membership degree array for $i^{th}$ service provider. The evaluation method of service providers is presented in (2).

$$V_i = (\alpha)TRV_i + (1-\alpha)SP_i^{Good}. \quad (2)$$
$$0 \leq \alpha \leq 1; \; 1 \leq i \leq n$$

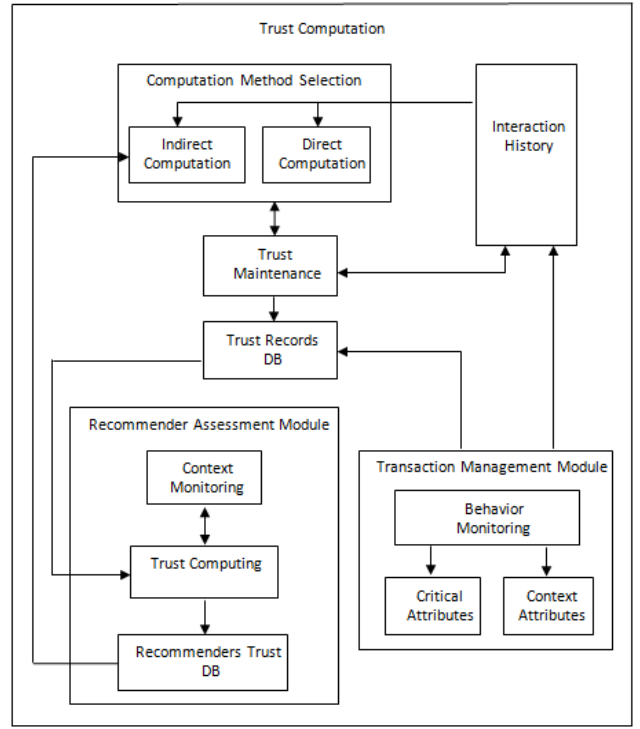

Figure 4. Trust Computation Module

where $V_i$ is the result value for the evaluation of $i^{th}$ service provider, *n* is the number of service providers, $SP_i^{Good}$ represents the membership degree of $i^{th}$ service provider with respect to the quality level of *Good*, $TRV_i$ is the trust value for $i^{th}$ service provider, and α is the weight factor which denotes the importance of the trust value in the computation. Equation (3) shows the selection method of the target entity.

$$TE = (entity_i \mid V_i = \max(V_z) \text{ and } 1 \leq i, z \leq k). \quad (3)$$

Here, *k* is the number of service providers and *TE* denotes the $i^{th}$ entity ($entity_i$) which has the maximum evaluation result value ($V_i$). Finally, $entity_i$ is selected as service provider.

TRUST COMPUTATION MODULE: Selecting an entity as a service provider depends on the trust value that the service requester makes on the entity. *Trust computation* module is responsible for computing trust values. *Trust computation* module helps *request management* module to select the service provider that has an acceptable trust value and plays a fundamental role in our trust management model.

IV. TRUST MANAGEMENT MODEL AT SERVICE PROVIDER

*Request responder* component responds to service requests and consists of *request processor* module and *privacy agent*.

```
Begin
  For all Att_i in the requested service where 1 ≤ i ≤ n
  If (RV_i ≥ TV_i)
      Service can be provided;
      Request responder component responds to the request management
        module with real values;
  Else
      Request processor module searches trust records DB;
      If (there is a trust record with an acceptable trust value for the
         requested service type)
          Request processor module recommends another entity
            corresponding to service provider ID of the trust record;
          Request responder component responds to the request
            management module with the trust record;
          The indirect trust value is computed by indirect trust
            computation method;
      Else
          Request responder component does not respond to request
            management module;
      End if
  End if
End
```

Figure 5. The request processing function

PRIVACY AGENT: Each entity in the pervasive environment has an agent (*privacy agent*) to maintain its security and privacy policies. These policies restrict irregular accesses and do not allow some services to be used by other entities. Furthermore, some entities do not like others to be aware of their context (e.g. location). *Privacy agent* is composed of *local policy* module and *context assessment* module described below.

LOCAL POLICY MODULE: This module is responsible for authentication. It also specifies access levels and access rules by considering the defined policy-levels and decides whether to forward the request to *request processor* module or to reject it.

CONTEXT ASSESSMENT MODULE: This module evaluates service attributes and critical contexts, and assigns each of them a privacy-level. Privacy-levels influence the decisions made by *local policy* module.

REQUEST PROCESSOR MODULE: The request which is passed from *local policy* module arrives to *request processor* module. *Request processor* module determines whether the corresponding entity can provide the service attributes with the values greater than the thresholds or not. In the former case, *request responder* component responds to the *request management* module at service requester with real values. In the latter case, *request processor* module searches *trust records DB* to find another entity which can provide the service and has a good trust value. The entity is then recommended to the service requester. Fig. 5 shows the request processing function where $n$ is the number of service attributes, $RV_i$ is the real value, and $TV_i$ is the threshold value for $i^{th}$ attribute.

V. TRUST COMPUTATION IN THE PROPOSED MODEL

As shown in Fig. 4, the following modules are responsible for trust computing.

TRUST RECORDS DB: It is a repository consisting of trust records. The important fields of a record are: service type, service attributes, last updated time and service provider ID.

TRUST MAINTENANCE: This module initializes, fetches, and updates records in *trust records DB*. If the last updated time field of a record contains an expired time, then it has to be updated.

INTERACTION HISTORY: It is a repository consisting of records that each record contains service attributes, context attributes, satisfaction degree, and the interaction time. For each interaction there exists a record in *interaction history*.

TRANSACTION MANAGEMENT MODULE: This module monitors the behavior of each transaction and then calculates the satisfaction degree. Context and critical attributes directly influence on satisfaction degree. Satisfaction degree is computed as in (4).

$$SD = \sum |EV_i^{norm} - PV_i^{norm}|/n. \qquad (4)$$
$$1 \le i \le n$$

where *SD* is the satisfaction degree, $n$ is the number of attributes, $EV_i^{norm}$ is the expected value, and $PV_i^{norm}$ is the provided value for the attributes that are normalized.

COMPUTATION METHOD SELECTION: In the case that there is not any trust for a specific entity in *trust records DB*, or the *trust records DB* needs to be updated, this module computes the trust value by selecting the corresponding computation method. In the case that there are adequate records in *interaction history* and the occurrence times are acceptable, trust value is computed directly. Otherwise, trust value is computed indirectly by the help of recommenders as shown in Fig. 6.

Trust computations are mostly based on the records in *interaction history*. Therefore, it is important to treat each record in *interaction history*, considering the interaction time. The results of recent interactions, which represent the current behavior of the entity, are more important than those of older interactions. Hence, we give weights to records based on the time they occur. Equation (5) computes the direct trust value for an entity.

$$DT = \sum ((W)^{(t^{cur}-t_i^{occ})} \cdot SD_i) / \sum (W)^{(t^{cur}-t_i^{occ})}. \qquad (5)$$
$$1 \le i \le k\ ;\ 0 < W \le 1$$

where *DT* represents the direct trust value, $SD_i$ represents the satisfaction degree for $i^{th}$ interaction, $t^{cur}$ is the current time, $t_i^{occ}$ is the occurance time of the $i^{th}$ interaction, $W$ is a weight factor which is used to give a moving weight $((W)^{(t^{cur}-t_i^{occ})})$ to $i^{th}$ interaction based on the occurance time, and $k$ is the number of interactions with the corresponding entity.

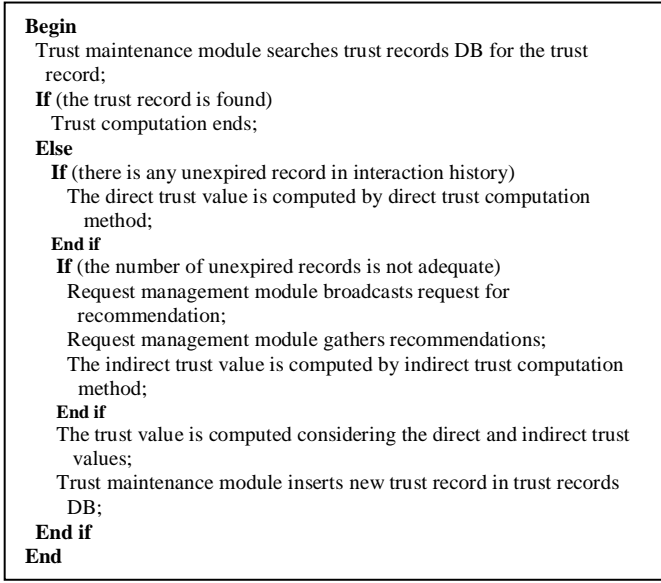

```
Begin
  Trust maintenance module searches trust records DB for the trust
    record;
  If (the trust record is found)
    Trust computation ends;
  Else
    If (there is any unexpired record in interaction history)
      The direct trust value is computed by direct trust computation
        method;
    End if
    If (the number of unexpired records is not adequate)
      Request management module broadcasts request for
        recommendation;
      Request management module gathers recommendations;
      The indirect trust value is computed by indirect trust computation
        method;
    End if
    The trust value is computed considering the direct and indirect trust
      values;
    Trust maintenance module inserts new trust record in trust records
      DB;
  End if
End
```

Figure 6. The process of trust computation

RECOMMENDER ASSESSMENT MODULE: Different recommenders have different weights that can be mentioned as their trust values. *Recommender assessment* module judges recommenders according to their honesty and context. Recommenders which are more trustworthy have more effect in computing the trust value of the recommended entity.

Equation (6) represents the initialization function of a recommender's trust value.

$$RT = \sum_{0 \leq i \leq n} TRV_i / n. \qquad (6)$$

where $RT$ is recommender's trust value, $n$ is the number of records in *trust records DB* that their service provider ID is same as the recommender ID, and $TRV_i$ is the trust value of the $i^{th}$ trust record. Recommenders will be updated after each interaction with the corresponding recommended entity. The similarity distance between the provided value and the real value is computed as in (7).

$$\delta = \sum_{1 \leq i \leq n} | PV_i^{norm} - RV_i^{norm} | / n. \qquad (7)$$

where $\delta$ represents the similarity distance, the shorter the distance means the more accurate recommender, $PV_i^{norm}$ is the normalized provided value and $RV_i^{norm}$ is the normalized recommended value for $i^{th}$ attribute, and $n$ is the number of attributes. The update factor for a recommender's trust value is computed as in (8).

$$UF = 1 - (\delta / error^{acpt}). \qquad (8)$$

where $error^{acpt}$ is the acceptable error between provided and recommended values and $UF$ is the update factor for the recommender's trust value. Finally, a recommender's trust value is updated as in (9).

$$RT^{new} = \begin{cases} +1 & if\,(1+UF) \cdot RT^{old} \geq +1 \\ -1 & if\,(1+UF) \cdot RT^{old} \leq -1 \\ (1+UF) \cdot RT^{old} & Otherwise \end{cases} \cdot \qquad (9)$$

where $RT^{old}$ is the old and $RT^{new}$ is the updated value for a recommender trust and $UF$ is the update factor. A recommender's trust value will be increased in the case of having less error. An unaccepted error causes the recommender trust value to be decreased.

As shown in Fig. 4, *recommender assessment* module is responsible for context monitoring. The context of a recommender can effect directly on the recommender's trust value. *Recommender assessment* module uses a rule-based evaluation method to evaluate the context of a recommender. The evaluation decreases the recommender's trust value in the case of unsuitable contexts. For example, the trust value of a long distance recommender is decreased according to special service types.

The indirect trust value for a recommended entity is computed as in (10).

$$IT = \sum_{1 \leq i \leq n} (RT_i \cdot TRV_i) / \sum RT_i. \qquad (10)$$

where $IT$ is the indirect trust value for the recommended entity, $n$ is the number of recommenders for that recommended entity, $RT_i$ is the recommender's trust value corresponding to $i^{th}$ recommender, and $TRV_i$ is the trust value which is recommended by the $i^{th}$ recommender.

The trust value for a service provider is computed according to the direct and indirect trust values which are described previously. The trust value affects *request management* module directly on the selection of a service provider. Equation (11) computes the trust value for an entity.

$$TRV = (\beta)DT + (1-\beta)IT. \qquad (11)$$
$$0 \leq \beta \leq 1$$

where $TRV$ is the trust value, $DT$ is the direct trust value, $IT$ is the indirect trust value, and $\beta$ is a factor which gives weight to the direct and indirect trust values. In the case that there exist enough unexpired interaction records in *interaction history*, $\beta$ is equal to 1.

## VI. Characteristics of the Proposed Model

A pervasive computing environment has a dynamic nature. Therefore, new entities constantly enter the environment. A new entity is unknown to all other entities in the environment No recommendation is available for the new entity. In this case, it is important to determine how the new entity can build trust relationship with other entities. In the proposed model, we assign the trust value of zero to the new entity. Thus, the interactions with the new entity can happen when other entities have negative trust values (untrustworthy entities).

Recommendations help a service requester to compute an indirect trust in the case that there are not adequate records in *interaction history* for direct trust computation. False recommendations effect on the computed trust value. Dishonest and malicious recommenders can provide false recommendations. In the proposed model, dishonest recommenders are identified and all recommendations provided by dishonest recommenders are excluded from indirect trust computation. To identify a dishonest recommender, the service requester uses all recommendations which are received from a specific recommender and computes the mean value of the recommended trust values. In the case that the mean value is so low or so high (not in an adequate range), the service requester judges the recommender to be dishonest.

The method of assigning weights to the interactions over time causes each past interaction to be effective in trust computing according to the assigned weight. Therefore, the weighting mechanism can protect the entity against the dynamic behavior of malicious recommenders.

*Context aware agent* in the trust management model provides a service selection mechanism which is based on contexts. As a result, target entities are restricted to the domains which are identified by *context aware agent*. Sending requests to domains, considering the context, facilitates the functionality of *request management* module and in this case, service providers with accurate context have more priority over other service providers.

## VII. Conclusion and Future Works

In this paper, we proposed a trust management model for pervasive computing systems based on the concept of context. We specified the details of each of the main components, and we presented the adjustments to the proposed methods that are needed to make the trust computation more accurate. Because of the fuzzy and dynamic nature of trust, we considered fuzzy concepts in our model. The trust value of an entity is dynamically updated after each related interaction. We provided an acceptable privacy-level in our trust management model to handle security and privacy protection of the pervasive computing.

It is important to provide an adaptive trust mechanism which safeguards service interactions in the dynamic and uncertain pervasive environment. The structure of the proposed models is needed to make the actual applications feasible. In the future we are going to work on implementing and simulating adaptive trust management models for dynamic and uncertain environments.

Acknowledgment

This work was supported by Iran Telecommunication Research Center (ITRC).